\documentclass[11pt]{article}
\usepackage{amsmath}
\usepackage{amsfonts}
\usepackage{amssymb, amsmath, cite}

\newtheorem{theorem}{Theorem}
\newtheorem{lemma}{Lemma}

\newtheorem{remark}{Remark}

\newcommand\be{\begin{equation}}
\newcommand\ee{\end{equation}}
\newcommand\ber{\begin{eqnarray}}
\newcommand\eer{\end{eqnarray}}
\newcommand\berr{\begin{eqnarray*}}
\newcommand\eerr{\end{eqnarray*}}

 \newcommand\re{\mathrm{e}}
  \newcommand\ri{\mathrm{i}}
\newcommand{\ud}{\mathrm{d}}
\newcommand{\nm}{\nonumber}

\newcommand{\ito}{\int_{\Omega}}

\newcommand{\vep}{\varepsilon}

\title{\bf Existence of Doubly  Periodic Vortices in  a Generalized Chern--Simons  Model\footnote{This work was supported by the National
Natural Science Foundation of China grant 11026077} }
\author{Xiaosen Han\footnote{Email: xiaosenhan@gmail.com}
\\ {\small \em{ Institute of Contemporary Mathematics, Henan University, Kaifeng 475004, China}}
\\ {\small\em{ School of Mathematics and Information Science, Henan University, Kaifeng 475004, China}}}
\date{}

\begin{document}
\maketitle
\begin{quote}{
{{\bfseries Abstract.}
 We establish an  existence  theorem for  the doubly periodic vortices in  a generalized self-dual Chern--Simons
 model. We show that there exists a critical value of the
 coupling parameter such that there exits self-dual doubly periodic vortex solutions for the  generalized self-dual Chern--Simons
 equation if and only if the coupling
 parameter is  less than or equal to the value.  The  energy, magnetic flux, and electric charge associated to the
 field configurations are all specifically quantized. By
 the  solutions obtained for this  generalized self-dual Chern--Simons
 equation we can also construct doubly periodic vortex solutions to a
 generalized self-dual Abelian Higgs  equation.}}

\end{quote}

\section{Introduction}
\setcounter{equation}{0}

 Vortices, which arise in   spontaneous broken gange theories in two-space dimension, play important roles in many areas of
physics including    superconductivity \cite{abri,jata,gila},
 optics\cite{bec}, cosmology\cite{hiki,kibb,vish},  and the quantum Hall
effect \cite{soko}. In recent years much attention has been devoted
to the study of  vortices  in $(2+1)$-dimensional Chern--Simons
gauge theory. An important feature of such vortices is that  they
are both magnetically and electrically charged, which is different
from the neutral Nielsen--Olesen vortices \cite{niol}. In the work
of Hong, Kim, and Pac\cite{hkp} and Jackiw and Weinberg\cite{jw1},
the Yang--Mills (or Maxwell) term is removed from the action
Lagrangian density while the Chern--Simons term alone governs
electromagnetism, which is physically sensible at large distances
and low energies. When the Higgs potential takes a special form as
that in the classical Abelian Higgs model\cite{bogo,jata}, the
static equations of motion can be reduced from a second-order
differential to a Bogolmol'ny type (self-dual) system of first-order
equations \cite{bogo}, which
 enables   one to make rigorous mathematical studies of such solutions.
In such a setting, topological multivortices with quantized
charges\cite{wangr,spya4}, non-topological multivortics with
fractional values of charges \cite{spya3,chmy,chim,cfl} and doubly
periodic vortices with quantized charges
\cite{caya1,taran96,cho1,djlw,linyan,tara5,nota1} are all present.

In \cite{bct}  Burzlaff, Chakrabarti, and  Tchrakian proposed a
generalized self-dual Chern--Simons--Higgs model and a generalized
Abelian Higgs model.  The non-topological  and topological vortices
for  the models  were  established more than ten years ago in
\cite{tcya} and \cite{yls1}, respectively. However, up to now, the
existence of  doubly periodic vortices  for the models  is still
open. Our purpose of this paper is to establish the existence of
doubly periodic multivortices  to the generalized  self-dual
Chern--Simons  model.
 As in \cite{tcya,yls1} we first  reduce  the generalized self-dual Chern--Simons
 equations  into a scalar quasilinear elliptic equation  with Dirac source terms characterizing  the locations of the vortices.
 Then by a transformation the  quasilinear elliptic equation can be
 simplified further  into a semilinear one.
We establish an existence theorem by  applying  a sub-super solution
method, which was used  by Caffarelli and Yang\cite{caya1} to
construct multiple doubly periodic vortex solutions to the
Chern--Simons model proposed in \cite{hkp,jw1}.

 The rest of our paper is organized as follows. In section 2,
we formulate our problem  and state our main results. In section 3
we prove the existence of doubly periodic vortices for the
generalized self-dual Chern--Simons equation. In section 4 we
construct a doubly periodic vortex solution for the generalized
self-dual Abelian Higgs model using our results in the previous
section.

\section{Generalized Chern--Simons vortices}\label{s2}

\setcounter{equation}{0}

We consider the generalized self-dual Chern--Simons equations
derived in \cite{bct} over a doubly periodic domain $\Omega$ such
that the field configurations are subject to the 't Hooft boundary
condition \cite{hoof,waya,yang1} under which periodicity is achieved
modulo gauge transformations.

Following \cite{bct}, we  derive  the generalized self-dual
Chern--Simons equations.
 The Lagrangian density in $(2+1)$ dimensions reads
  \be \mathcal {L}= \sqrt2\kappa\vep^{\mu\nu\alpha}\left[A_\alpha-2\ri\left(1-\frac12|\phi|^2\right)\phi
  \overline{D_\mu\phi}\right]F_{\mu\nu}+2(1-|\phi|^2)^2|D_\mu\phi|^2-V, \label{a0}\ee
where $D_\mu=\partial_\mu+\ri A_\mu$ is the gauge-covariant
derivative, $A_\mu \, (\mu=0,1,2)$   a 3-vector gauge field, $\phi$
  a complex scalar field called the Higgs field,
$F_{\alpha\beta}=\partial_\alpha A_\beta-\partial_\beta A_\alpha$
 the induced electromagnetic field,  $\alpha, \beta, \mu,\nu=0,
1,2$, $\kappa>0$ is a
  constant referred to as the Chern--Simons
  coupling parameter, $\vep^{\alpha\beta\gamma}$ the Levi-Civita totally
  skew-symmetric tensor with $\vep^{012}=1$,  $V$  the Higgs
  potential function, and the summation convention over repeated indices
  is observed.

Varying \eqref{a0} with respect to $A_\alpha$ and $\phi$,  we have
the Euler-Lagrange equations
 \ber
 &&\frac{1}{2\sqrt2}\kappa\vep^{\mu\nu\alpha}\left[(1-|\phi|^2)F_{\mu\nu}-\ri\left(D_\mu\phi \overline{D_\nu\phi}-D_\nu\phi \overline{D_\mu\phi}\right)\right]
 +\ri(1-|\phi|^2)\phi\overline{D^\alpha\phi}=0, \label{aa1}\\
  &&2\sqrt2\kappa\ri\vep^{\mu\nu}D_{\mu}\left[\left(1-\frac12|\phi|^2\right)\phi\right]F_{\mu\nu}+2\partial^\mu\left[\left(1-|\phi|^2\right)^2\right]D_\mu\phi
   \nm\\&&\qquad+2\left(1-|\phi|^2\right)^2D^\mu D_\mu\phi-4\left(1-|\phi|^2\right)\phi|D_\mu\phi|^2-\frac{\partial V}{\partial
  \overline{\phi}}=0.\label{aa1'}
 \eer
 In the static limit,  the $\alpha=0$ component of \eqref{aa1}
 implies
  \be A_0=-\frac{\kappa}{\sqrt2|\phi|^2\big(1-|\phi|^2\big)}\left[\big(1-|\phi|^2\big)F_{12}
  -\ri \big(D_1\phi\overline{D_2\phi}-\overline{D_1\phi}D_2\phi\big)\right]. \label{aa2}
  \ee
From \eqref{aa2}, we can express the density of electric  charge as
 \be
  \rho=-\sqrt2A_0|\phi|^2\big(1-|\phi|^2\big)=\kappa\left[\big(1-|\phi|^2\big)F_{12}
  -\ri \big(D_1\phi\overline{D_2\phi}-\overline{D_1\phi}D_2\phi\big)\right].\label{aa3}
 \ee

 Note that the  energy $E$ can be expressed as
 \berr
  E=\ito\ud
  x\left[2\big(1-|\phi|^2\big)^2\big(|D_0\phi|^2+|D_1\phi|^2+|D_2\phi|^2\big)+V\right].
  \eerr
With the choice of  the  Higgs potential
   \[V=\frac{1}{4\kappa^2}(1-|\phi|^2)^4|\phi|^2, \]  in the static limit we have
 \berr
  E=\ito\ud x\left\{\frac{\kappa^2}{|\phi|^2}\left[\big(1-|\phi|^2\big)F_{12}-\ri \big(D_1\phi\overline{D_2\phi}-\overline{D_1\phi}D_2\phi\big)\right]^2
  \right.\\
  \left.+2\big(1-|\phi|^2\big)^2\big(|D_1\phi|^2+|D_2\phi|^2\big)+\frac{1}{4\kappa^2}|\phi|^2\big(1-|\phi|^2\big)^4 \right\}.
 \eerr
Then we rewrite  the energy as
 \berr
 E&=&\ito\ud x\left\{\left( \frac{\kappa}{|\phi|}\left[\big(1-|\phi|^2\big)F_{12}-\ri \big(D_1\phi\overline{D_2\phi}-\overline{D_1\phi}D_2\phi\big)\right]
 -\frac{1}{2\kappa}|\phi|\big(1-|\phi|^2\big)^2 \right)^2\right.\\
 &&\left. +2\big(1-|\phi|^2\big)^2|D_1\phi-\ri D_2\phi|^2+\big(1-|\phi|^2\big)^3F_{12}-3\ri\big(1-|\phi|^2\big)^2\big(D_1\phi\overline{D_2\phi}-\overline{D_1\phi}D_2\phi\big)\right\}
\\
 &=&\ito\ud x\left\{\left( \frac{\kappa}{|\phi|}\left[\big(1-|\phi|^2\big)F_{12}-\ri \big(D_1\phi\overline{D_2\phi}-\overline{D_1\phi}D_2\phi\big)\right]
 -\frac{1}{2\kappa}|\phi|\big(1-|\phi|^2\big)^2 \right)^2\right.\\
 &&\left.+2\big(1-|\phi|^2\big)^2|D_1\phi-\ri D_2\phi|^2+ F_{12}-3\ri\vep_{ij}\partial_i\left[\left(1-|\phi|^2+\frac13|\phi|^4\right)\phi\overline{D_j\phi}\right]\right\}.
 \eerr
Consequently, we have
 \be
 E\ge \ito F_{12}\ud x, \label{aa4}
 \ee
 and the lower bound is saturated if and only if  $(\phi, A)$ satisfies  the self-dual
 equations
  \ber
  D_1\phi&=&\ri D_2\phi, \label{a1}\\
   \big(1-|\phi|^2\big)F_{12}&=&\ri \big(D_1\phi\overline{D_2\phi}-\overline{D_1\phi}D_2\phi\big)+\frac{1}{2\kappa^2}|\phi|^2\big(1-|\phi|^2\big)^2. \label{a2}
  \eer

    We aim to seek doubly
periodic  $N$-vortex solutions of \eqref{a1} and \eqref{a2} such
that, $\phi$ vanishes at the arbitrarily prescribed points, $p_1,
p_2, \dots, p_m\in \Omega$ with multiplicities $n_1, n_2, \dots,
n_m$, repectively, and $\sum_{i=1}^m n_i=N$.

Our main result for the existence of periodic multiple vortices of
\eqref{a1} and \eqref{a2} reads as follows.

  \begin{theorem} \label{th1}
   Let $p_1, p_2, \dots, p_m\in\Omega$, $n_1, n_2, \dots, n_m$ be
   some  positive integers and $N=\sum_{i=1}^mn_i$. There exists a critical
   value of the coupling parameter, say $\kappa_c$, satisfying
   \[0<\kappa_c \le  \sqrt{\frac{|\Omega|}{27\pi N}},\]
   such that the self-dual equations
 \eqref{a1} and \eqref{a2} admit a solution $(\phi, A)$ for which $p_1,
 p_2,\dots,p_m$ are zeros of $\phi$ with multiplicities $n_1, n_2, \dots,
 n_m$,   if and only if $0<\kappa\le \kappa_c$.

  The solution $(\phi, A)$ also satisfies the following
   properties.

    The energy, magnetic flux, and electric charge  are given by
    \be
     E=2\pi N,\quad \Phi=2\pi N, \quad Q=2\kappa\pi N.\label{a3}
    \ee

    The solution  $(\phi, A)$ can be chosen such that the magnitude
    of $\phi$, $|\phi|$ has the largest possible values.

     Let the prescribed data be denoted by $S=\{p_1, p_2,\dots p_m; n_1, n_2,
    \dots,n_m\}$, where $n_i$ may be zero for $i=1,\dots,m$, and
    denote the dependence of $\kappa_c$ on $S$ by $\kappa_c(S)$. For
    $S'=\{p_1, p_2,\dots p_m; n'_1, n'_2, \dots,\\n'_m\}$, we denote $S\le S'$
    if $n_1\le n_1', \dots, n_m\le n_m'$. Then $\kappa_c$ is a
    decreasing function of $S$ in the sense that
     \be \kappa_c(S)\ge\kappa_c(S'), \quad {if}\quad S\le S'.\label{a4}\ee
  \end{theorem}

\section{ Existence of doubly periodic vortices}\label{s3}

\setcounter{equation}{0} \setcounter{theorem}{0}

 Following \cite{yls1}, we first  rewrite the equations \eqref{a1} and
 \eqref{a2} as a quasilinear elliptic equation with the Dirac source
 terms.

Using \eqref{a1}, we have
\[\ri \big(D_1\phi\overline{D_2\phi}-\overline{D_1\phi}D_2\phi\big)= -\big(|D_1\phi|^2+|D_2\phi|^2\big)\]
  Then we can  rewrite \eqref{a2} in the form
  \be
  \big(1-|\phi|^2\big)F_{12}=-\big(|D_1\phi|^2+|D_2\phi|^2\big)+\frac{1}{2\kappa^2}|\phi|^2\big(1-|\phi|^2\big)^2.
  \label{b1}
  \ee
We complexify the variables
 \[z=x^1+\mathrm{i}x^2,  \quad A=A_1+\mathrm{i}A_2.
 \]
Let
\[\partial=\frac12(\partial_1-\mathrm{i}\partial_2)\quad \bar{\partial}=\frac12(\partial_1+\mathrm{i}\partial_2).\]
 Then by  \eqref{a1}, we can get, away from the zeros of $\phi$,
 \be
  F_{12}=-2\partial\bar{\partial}\ln|\phi|^2=-\frac12\Delta\ln|\phi|^2.\label{b2}
  \ee
Introduce  the real  variable $u=\ln |\phi|^2$. A direct computation
leads to
 \be
 |D_1\phi|^2+|D_2\phi|^2=\frac12\re^u|\nabla u|^2.\label{b3}
 \ee

Counting all  the multiplicities of the zeros of $\phi$, we write
 the prescribed zero set as $Z(\phi)=\{p_1, \dots, p_N\}$.
Inserting \eqref{b3} into \eqref{b2}, the equations \eqref{a1} and
\eqref{a2} are transformed into the  following quasilinear elliptic
equation
 \be
  (1-\re^u)\Delta u-\re^u|\nabla u|^2=-\lambda \re^u(\re^u-1)^2+4\pi\sum\limits_{s=1}^N\delta_{p_s} \quad \text{in }\quad \Omega,  \label{b4}
 \ee
where  \[\lambda=\frac{1}{\kappa^2},\] and  $\delta_p$ is the Dirac
distribution centered at $p\in\Omega$.

Conversely, if $u$ is a solution of \eqref{b4}, we can obtain a
solution of \eqref{a1}-\eqref{a2} according to the transformation
 \ber
  &&\phi(z)=\exp{\left(\frac12u(z)+\mathrm{i}\sum\limits_{s=1}^N\arg(z-p_s)\right)},\label{b4'}\\
   &&A_1(z)=-2\mathrm{Re}\{\mathrm{i}\bar{\partial}\ln\phi\}, \quad A_2(z)=-2\mathrm{Im}\{\mathrm{i}\bar{\partial}\ln\phi\}.\label{b4''}
  \eer
Hence  it is sufficient to solve \eqref{b4}. Indeed we can establish
the following  existence result for \eqref{b4}.
\begin{theorem}\label{th2}
 For any prescribed  points $p_1, \dots, p_N\in\Omega$,  there is a critical value of $\lambda$, say  $\lambda_c$, satisfying
 \[
  \lambda_c\ge \frac{27\pi N}{|\Omega|},
 \]
  such that, the equation \eqref{b4} has a negative solution if and only if  $\lambda\ge\lambda_c$. Moreover, there holds
  the quantized integral
   \be
   \lambda\ito\re^u\big(\re^u-1\big)^2\ud x=4\pi N. \label{b7'}
   \ee

   Let the prescribed data be denoted by $S=\{p_1, p_2,\dots p_m; n_1, n_2,
    \dots,n_m\}$, where $n_i$ may be zero for $i=1,\dots,m$, and
    denote the dependence of $\lambda_c$ on $S$ by $\lambda_c(S)$. For
    $S'=\{p_1, p_2,\dots p_m; n'_1, n'_2, \dots, \\n'_m\}$, we denote $S\le S'$
    if $n_1\le n_1', \dots, n_m\le n_m'$. Then $\lambda_c$ is an
    increasing function of $S$ in the sense that
     \be \lambda_c(S)\le\lambda_c(S'), \quad {if}\quad S\le S'.\label{a4}\ee
\end{theorem}

By Theorem \ref{th2}, to complete the proof of Theorem  \ref{th1},
we just need to compute the energy, magnetic flux and electric
charge associated to the field configurations $(\phi, A)$.
  Let $u$ be a solution of \eqref{b4} obtained  in Theorem
  \ref{th2}. Then $(\phi, A)$ defined  by   \eqref{b4'} and
  \eqref{b4''}  is a $N$-vortex solution of \eqref{a1} and \eqref{a2}.

   By  \eqref{aa4}, \eqref{a1}, and \eqref{a2},  we have
    \ber
  E=\Phi=\ito F_{12}\ud x=-\frac12\ito\Delta u\ud x=-\frac12\lim\limits_{r\to0}\int_{\Omega\setminus\cup_{j=1}^NB_r(p_j)}\nabla \cdot\nabla u\ud
  x\nm\\
  =\frac12\sum\limits_{j=1}^N\lim\limits_{r\to0}\int_{ \partial B_r(p_j) }(-\partial_2u\ud x^1+\partial_1u\ud x^2). \label{b26}
  \eer
 where $B_r(p_j)$ is the disc in $\Omega$ centered at $p_j$ with
 radius $r>0 \, (j=1, \dots, N)$.

  Noting that near the the point $p_j$, we have the expression
   \be
    u(x)=\ln|x-p_j|^2+f_j(x), \quad f_j\in C^{\infty}(B_r(p_j)),\, j=1, \dots, N,\label{b24}
   \ee
where $r>0$ is small. Then, plugging \eqref{b24} into \eqref{b26},
we can obtain
 \be
  E=\Phi=2\pi N.\label{b25}
 \ee

From  \eqref{aa3},  the density  of the electric charge can be
expressed as
  \[\rho= \kappa\left[\left(1-|\phi|^2\right)F_{12}+\big(|D_1\phi|^2+|D_2\phi|^2\big)\right]=-\frac\kappa2(1-\re^u)\Delta u+\frac\kappa2\re^u|\nabla u|^2.\]
Therefore, by \eqref{b24},  the electric charge is
  \ber
  Q&=&\ito \rho\ud x\nm\\
  &=&-\frac\kappa2\ito\nabla\cdot[(1-\re^u)\nabla u] \ud x\nm\\
  &=&\frac\kappa2\sum\limits_{j=1}^N\lim\limits_{r\to0}\int_{ \partial B_r(p_j) } (1-\re^u)(-\partial_2u\ud x^1+\partial_1u\ud x^2)=2\kappa\pi N. \label{b27}
  \eer
From \eqref{b25} and \eqref{b27},  we obtain \eqref{a3}, which says
that the energy, magnetic flux,  and electric charge are all
quantized.

In what follows we only need to prove Theorem \ref{th2}.  To
simplify the problem further, we first derive  an  a priori estimate
 for the solutions of \eqref{b4}.
\begin{lemma}\label{le1}
 If $u$ is a solution to \eqref{b4}, then $u$ is negative throughout
 $\Omega$.
\end{lemma}

 {\bf Proof.}  Denote  \[B_\vep(p_j)=\{x|x\in \Omega, \, |x-p_j|<\vep\},\] and
 \be
 \Omega_\vep=\Omega\setminus \bigcup_{j=1}^N B_\vep(p_j ).\label{b3'}
 \ee
We see that $u$ is negative on $\partial\Omega_\vep$ when $\vep$ is
sufficiently small. Noting that
 \[ (1-\re^u)\Delta u+\lambda \re^u(\re^u-1)^2= \re^u|\nabla u|^2\ge0 \quad \text{in }\quad \Omega\setminus\{p_1, \dots, p_N\},\]
by the  maximum principle, we obtain  $u<0$ in $ \Omega_\vep$. Then
we have $u<0$ in $ \Omega$.

Then by Lemma \ref{le1}, to solve \eqref{b4},  we just need to
consider the negative solutions to \eqref{b4}.

 Since \eqref{b4} is quasilinear, it is difficult to deal with
 directly. Therefore, as in \cite{yls1,tcya}, we consider a new
 dependent variable $v$ defined by
 \be
  v=F(u)=1+u-\re^u.\label{b5}
  \ee
It is easy to see that $F'(t)=1-\re^t$, $F''(t)=-\re^t<0$. Then
$F(\cdot)$ is increasing and invertible over $(-\infty, 0]$.
Denoting the inverse of $F$ over $(-\infty, 0]$ by $G$, we see that
both $F$ and $G$ are $1\sim1$ from  $(-\infty, 0]$ to itself.

 In view of  the fact that  solutions to the equation \eqref{b4}  are all negative,
 the equation \eqref{b4}  is equivalent to the following semilinear
equation,
 \be
  \Delta v=-\lambda\re^{G(v)}\big(\re^{G(v)}-1\big)^2+4\pi\sum\limits_{j=1}^N\delta_{p_j} \quad \text{in }\quad
  \Omega.\label{b6}
 \ee
Then we just need to seek negative solutions to \eqref{b6}.

Let $v_0$ be a solution of the equation (see \cite{aubi})
 \be
 \Delta v_0=-\frac{4\pi N}{|\Omega|}+4\pi\sum\limits_{j=1}^{N}\delta_{p_j}. \label{b7}
 \ee
 Setting $v=v_0+w$,  then the equation \eqref{b6}  is  reduced to the
 following equation,
 \be
  \Delta w=-\lambda\re^{G(v_0+w)}\big(\re^{G(v_0+w)}-1\big)^2+ \frac{4\pi N}{|\Omega|} \quad \text{in }\quad
  \Omega.\label{b8}
 \ee

 In  the sequel we just need to consider \eqref{b8}.

 We easily  see that   the function $f(t)=-\re^t(\re^t-1)^2, t\in(-\infty, 0]$,  has a unique minimal value $-\frac{4}{27}$. If
$w$ is a solution of \eqref{b8}, then $v_0+w<0$.  Hence we  have
  \be
  \Delta w\ge-\frac{4}{27}\lambda +\frac{4\pi N}{|\Omega|}.\label{b9}
  \ee

   Then integrating
(\ref{b9}) over  $\Omega$,  we can obtain
  \be
  \lambda\ge\frac{27\pi N}{|\Omega|}, \label{b10}
 \ee
which is a necessary condition for the existence of solutions to
\eqref{b8}.

 As in \cite{caya1} or  Chapter 5 in \cite{yang1}  we can use a  super-sub solution method to establish the existence results for
\eqref{b8}.

It is easy to see that $w^*=-v_0$ is a supersolution to \eqref{b8}
in the distributional sense.

Then, in order  to solve \eqref{b8},  we introduce the following
iterative scheme
  \ber\left\{\begin{array}{lll}
  (\Delta-K)w_n=-\lambda\re^{G(v_0+w_{n-1})}\big(\re^{G(v_0+w_{n-1})}-1\big)^2-Kw_{n-1}+\frac{4\pi N}{|\Omega|} ,\\
  n=1,2,\dots,\\[1mm]
  w_0=-v_0,
\end{array}\label{b11}
\right.
  \eer
where $K$ is a  positive constant to be determined.

\begin{lemma}\label{lem1}
Let $\{w_n\}$ be  the sequence defined by \eqref{b11} with
$K>2\lambda$. Then \be
 w_0>w_1>w_2>\cdots>w_n>\cdots>w_*\label{b12}
 \ee
for any subsolution $w_*$ of \eqref{b8}. Therefore, if \eqref{b8}
has a subsolution,
 the sequence $\{w_n\}$ converge to a solution of \eqref{b8} in the space $C^k(\Omega)$
 for any $k\ge0$ and such a solution is the maximal solution of the equation \eqref{b8}.
\end{lemma}

{\bf Proof.} \quad We prove by \eqref{b12} by induction.

  When $n=1$, from \eqref{b11} we have,
  \[(\Delta-K) w_1=Kv_0+\frac{4\pi N}{|\Omega|}, \]
 which implies $w_1\in C^\infty(\Omega)\cap C^\alpha(\Omega)$ for some $0<\alpha<1$.
 Noting that
 \[(\Delta-K)(w_1-w_0)=0 \quad\text{in}\quad  \Omega\setminus\{p_1, p_2, \dots, p_N\},\]
 and $w_1-w_0<0$ on $\partial\Omega_\vep$, where $\Omega_\vep$ is
 defined by \eqref{b3'} for $\vep$ sufficiently small,  and using the maximum principle,  we have
$w_1-w_0<0$ in $\Omega_\vep$. Hence we  obtain $w_1<w_0 $ in
$\Omega$.

 Suppose  that $w_0>w_1>\cdots>w_k.$  It follows  from \eqref{b11} and
 $K>2\lambda$ that
 \berr
 && (\Delta-K)(w_{k+1}-w_k) \\
 &&=-\lambda\left[\re^{G(v_0+w_k)}\big(\re^{G(v_0+w_k)}-1\big)^2-\re^{G(v_0+w_{k-1})}\big(\re^{G(v_0+w_{k-1})}-1\big)^2\right]-K(w_k-w_{k-1})\\
  &&=\left[\lambda \re^{G(v_0+\xi)}\big(3\re^{G(v_0+\xi)}-1\big) -K\right](w_k-w_{k-1})\\
  &&\ge(2\lambda-K)(v_k-v_{k-1})\\
  &&\ge0,
 \eerr
 where  we have used the mean value theorem, $w_k\le \xi\le w_{k-1}$. Applying  the  maximum principle again, we  obtain $w_{k+1}<w_k$ in $\Omega$.

 Now we prove the lower bound in \eqref{b12} in terms of the subsolution $w_*$ of \eqref{b8}. That is,  $w_*\in C^2(\Omega)$ and
 \be
 \Delta w_*\ge-\lambda\re^{G(v_0+w_*)}\big(\re^{G(v_0+w_*)}-1\big)^2 +\frac{4\pi N}{|\Omega|}.\label{b13}
 \ee
Noting that $w_0=-v_0$ and \eqref{b13}, we have
 \berr
  \Delta(w_*-w_0)&\ge&-\lambda\re^{G(w_*-w_0)}\big(\re^{G(w_*-w_0)}-1\big)^2\\
   &=&2\lambda\re^{G(w_*-w_0)}\re^{G(\xi-w_0)}(w_*-w_0) \quad\text{in}\quad\Omega\setminus\{p_1,\dots, p_N\},
 \eerr
 where $\xi$ lies between $w_*$ and $w_0$.
 If $\vep>0$ is small, we see that $w_*-w_0<0$ on $\partial\Omega_\vep$. Then, by the maximum principle, we obtain $w_*-w_0<0$ in $\Omega_\vep$. Therefore, $w_*<w_0$ throughout $\Omega$.

Now assume $w_*<w_k$ for some $k\ge0$. It follows from \eqref{b11},
\eqref{b13},  and the fact  $K>2\lambda$ that
 \berr
  \Delta (w_*-w_{k+1})&\ge& -\lambda\left[\re^{G(v_0+w_*)}\big(\re^{G(v_0+w_*)}-1\big)^2-\re^{G(v_0+w_k)}\big(\re^{G(v_0+w_k)}-1\big)^2\right]-K(w_*-w_k)\\
  &=&\left[\lambda \re^{G(v_0+\xi)}\big(3\re^{G(v_0+\xi)}-1\big) -K\right](w_*-w_k)\\
  &\ge&(2\lambda-K)(w_*-w_k)\\
  &\ge&0,
 \eerr
where   $w_*\le \xi\le w_k$.  Using the maximum principle again, we
get $w_*<w_{k+1}$. Then we get \eqref{b12}.

Following  a standard bootstrap argument, we can obtain the
convergence of the sequence $\{v_n\}$ in any $C^k(\Omega)$.

In the sequel  we  only need to construct  a subsolution of
\eqref{b8}. Indeed, we can establish the following lemma.
\begin{lemma}\label{lem2}
 If $\lambda>0$ is sufficiently large,  the equation \eqref{b8} admits a subsolution  satisfying \eqref{b13}.
\end{lemma}

{\bf  Proof.} \quad  Take  $\vep>0$ sufficiently small such that the
balls
 \[ B(p_j, 2\vep)=\big\{x\in \Omega| \quad |x-p_j|<2\vep\big\}, \,\, j=1, 2, \cdots, N, \]
verify $B(p_i, 2\vep)\bigcap B(p_j, 2\vep)=\emptyset,$  if  $i\neq
j$.  Let $f_\vep$ be a smooth function defined on $\Omega$ such that
$0\le f_\vep\le 1$ and
  \berr f_\vep=\left\{\begin{array}{lll}1,&x\in B(p_j, \vep), \,\,  j=1, 2, \cdots, N,\\
  0, &x\notin\bigcup\limits_{j=1}^N B(p_j, 2\vep),\\
    \text{ smooth connection}, &\text  { elsewhere.}
\end{array}
\right.
  \eerr

  Then,
  \be
 \bar{ f}_\vep\equiv\frac{1}{|\Omega|}\ito f_\vep\ud x\le \frac{4\pi N\vep^2}{|\Omega|}. \label{b14}
  \ee
Define
\[g_\vep= \frac{8\pi N}{|\Omega|}(f_\vep-\bar{f}_\vep).\]
 It is easy to see that \[\ito g_\vep\ud x=0.\] Then we  see that the linear elliptic equation
 \be
  \Delta w=g_\vep  \quad \text{in}\quad \Omega,\label{b15}
 \ee
admits a unique solution up to an additive constant.

 When  $x\in B(p_j, \vep)\,(\, j=1, 2, \cdots, N)$, it follows from \eqref{b14} that
\be
 g_\vep\ge \frac{4\pi N}{|\Omega|}\left(2-\frac{8\pi N\vep^2}{|\Omega|}\right)> \frac{4\pi N}{|\Omega|},\label{b16}
\ee if $\vep$ is sufficiently  small. In the sequel we fix $\vep$
such that \eqref{b16} is valid.

   Now we choose a solution of \eqref{b15}, say  $\underline{w}$, to satisfy
   \[ v_0+\underline{w} \le 0, \,\, x\in \Omega.\]
Hence, for any $\lambda>0,$ we have
 \be
 \Delta \underline{w}=g_\vep> \frac{4\pi N}{|\Omega|}\ge-\lambda\re^{G(v_0+\underline{w})}\big(\re^{G(v_0+\underline{w})}-1\big)^2+ \frac{4\pi N}{|\Omega|},\label{b17}
 \ee
for $x\in B(p_j, \vep), \, j=1, 2, \cdots, N.$

 Finally, set
 \berr
  \mu_0=\inf\left\{\re^{G(v_0+\underline{w})} \left| x\in \Omega\setminus\bigcup\limits_{j=1}^N B(p_j, \vep)\right.\right\},\\
  \mu_1=\sup\left\{\re^{G(v_0+\underline{w})} \left| x\in \Omega\setminus\bigcup\limits_{j=1}^N B(p_j, \vep)\right.\right\}.
 \eerr
Then $0<\mu_0<\mu_1$ and

 \[
  -\re^{G(v_0+\underline{w})}\big(\re^{G(v_0+\underline{w})}-1\big)^2\le-\mu_0(1-\mu_1)^2\quad
  \text{for}\quad x\in \Omega\setminus\bigcup\limits_{j=1}^N B(p_j, \vep).
 \]
  Therefore, noting the boundedness  of $g_\vep$, we have
  \be
  \Delta \underline{w}=g_\vep \ge-\lambda\re^{G(v_0+\underline{w})}\big(\re^{G(v_0+\underline{w})}-1\big)^2
   +\frac{4\pi N}{|\Omega|}\quad \text{for}\quad x\in \Omega\setminus\bigcup\limits_{j=1}^N B(p_j, \vep),\label{b18}
  \ee
 if we take $\lambda$ large enough.

 Hence from \eqref{b17} and \eqref{b18} we infer that  $\underline{w}$ is
 a subsolution to \eqref{b8} if $\lambda$ is sufficiently large.

\begin{lemma}\label{lem3}
 There is a critical value of $\lambda$, say  $\lambda_c$, satisfying
 \be
  \lambda_c\ge \frac{27\pi N}{|\Omega|}, \label{b19}
 \ee
  such that, for $\lambda>\lambda_c,$ the equation \eqref{b8} has a solution, while for $\lambda<\lambda_c$,
  the equation  \eqref{b8} has no solution.
\end{lemma}

{\bf Proof.} \quad  Assume that $w$ is a solution of \eqref{b8}.
Then $v=v_0+w$ satisfies \eqref{b6} and is negative  throughout
$\Omega$.

 Define
 \[ \Lambda=\big\{\lambda>0\big| \lambda \,\,\text{is such that \eqref{b8} has a solution}\big\}.\]
Then we can prove  that  $\Lambda$ is an interval. To this end, we
prove that, if $\lambda'\in \Lambda$, then $[\lambda',
+\infty)\subset\Lambda$. Denote by $w'$ the solution of \eqref{b8}
at $\lambda=\lambda'$. Noting that $v_0+w'<0$, we see that $w'$ is a
subsolution of \eqref{b8} for any $\lambda>\lambda'$. By Lemma
\ref{lem1},  we obtain a solution of \eqref{b8} for any
$\lambda>\lambda'$. Hence $[\lambda', +\infty)\subset\Lambda$.

Set $\lambda_c=\inf\Lambda.$ Then, by the  \eqref{b10}, we have
$\lambda> \frac{27\pi N}{|\Omega|}$ for any $\lambda>\lambda_c$.
Taking the limit $\lambda\to\lambda_c$, we obtain \eqref{b19}.

  Let $w$ be a solution of \eqref{b8} we have just  obtained. Then $v=v_0+w$ is a solution to \eqref{b6} and $u=G(v)$ is a solution to \eqref{b4}.
  Hence, integrating  \eqref{b8} over $\Omega$, we have
 \[
 \lambda \ito \re^{G(v_0+w)}\big(\re^{G(v_0+w)}-1\big)^2\ud x=4\pi N,
 \]
which implies \eqref{b7'}.

Now we consider the critical case $\lambda=\lambda_c$. We use the
method of \cite{taran96} to deal with this. We first show that the
solution of  \eqref{b8} is monotonic with respect to $\lambda$.
\begin{lemma}\label{lem4}
 The maximum solutions of \eqref{b8}, $\{w_\lambda
|\lambda>\lambda_c\}$,  are a  monotone family in the sense that
$w_{\lambda_1}>w_{\lambda_2}$ whenever
$\lambda_1>\lambda_2>\lambda_c$.
\end{lemma}

 {\bf Proof.} \quad   Let $w_\lambda$ be a solution of \eqref{b8} obtained.  Then we have $u_0+w_\lambda<0$.  By the equation \eqref{b8}
 we obtain
 \berr
 \Delta w_{\lambda_2}&=&-\lambda_2\re^{G(v_0+w_{\lambda_2})}(\re^{G(v_0+w_{\lambda_2})}-1)^2+\frac{4\pi N}{|\Omega|}\\
 &=&-\lambda_1\re^{G(v_0+w_{\lambda_2})}(\re^{G(v_0+w_{\lambda_2})}-1)^2+\frac{4\pi N}{|\Omega|}+(\lambda_1-\lambda_2)\re^{G(v_0+w_{\lambda_2})}(\re^{G(v_0+w_{\lambda_2})}-1)^2\\
 &\ge&-\lambda_1\re^{G(v_0+w_{\lambda_2})}(\re^{G(v_0+w_{\lambda_2})}-1)^2+\frac{4\pi N}{|\Omega|}
 \eerr
for $\lambda_1>\lambda_2>\lambda_c$. Hence  $ w_{\lambda_2}$ is a
subsolution of \eqref{b8}  with $\lambda=\lambda_1$.   Then  by the
maximum principle,  we have  $w_{\lambda_1}>w_{\lambda_2}$ if
$\lambda_1>\lambda_2>\lambda_c$.

Next we show that solutions to \eqref{b8}  are all  bounded in
$W^{1,2}(\Omega)$.  We know that $W^{1,2}(\Omega)$ can be decomposed
as
  \[W^{1,2}(\Omega)=\mathbb{R}\oplus X.\]
  where
\[X=\left\{v\in W^{1,2}(\Omega)\Bigg| \ito v\ud x=0\right\}\]
  is a closed subspace of $W^{1,2}(\Omega)$.
In other words, for any $v\in W^{1,2}(\Omega)$, there exits a unique
number $c\in \mathbb{R}$ and $v'\in X$ such that $v=c+v'$.

\begin{lemma}\label{lem5}
 Let $w_\lambda$ be a solution of \eqref{b8}. Then $w_\lambda=c_\lambda+w'_\lambda$,
 where  $c_\lambda\in \mathbb{R}$ and $w'_\lambda\in X$.  We
 \be
 \|\nabla w'_\lambda\|_2\le C\lambda,\label{b.14}
 \ee
 where $C$ is a positive constant depending only on the size of the
 domain  $\Omega$. Furthermore, $\{c_\lambda\}$ satisfies the estimate
 \be|c_\lambda|\le C(1+\lambda+\lambda^2).\label{b.15}\ee
 Especially, $w_\lambda$ satisfies
 \be\|w_\lambda\|_{W^{1,2}(\Omega)}\le C(1+\lambda+\lambda^2).\label{b.16}\ee
\end{lemma}

{\bf  Proof.}\quad
 Noting  that  \be v_0+w_\lambda=v_0+c_\lambda+w'_\lambda<0, \label{b.01}\ee
 then multiplying the equation  \eqref{b8} by
$v'_\lambda$, integrating over $\Omega$, using the H\"{o}lder
inequality and the  Poincar\'{e} inequality, we  can obtain
 \berr
 \|\nabla w'_\lambda\|_2^2&=& \lambda\ito\re^{G(v_0+w_{\lambda})}(\re^{G(v_0+w_{\lambda})}-1)^2 w'_\lambda\ud x\\
 &\le&\lambda\ito|w'_\lambda|\ud x\le
 \lambda|\Omega|^{1/2}\|w'_\lambda\|_2\le C\lambda\|\nabla w'_\lambda\|_2,
 \eerr
which implies \eqref{b.14}.

 Using \eqref{b.01} again, we get  an  upper bound for  $c_\lambda$,
 \be c_\lambda<-\frac{1}{|\Omega|}\ito v_0(x)\ud x.\label{b.17}\ee

 Now we show that $c_\lambda$ is also bounded from below.  In view of \eqref{b.01}, it follows from the equation \eqref{b8} that
\[\Delta w_\lambda+\lambda\re^{G(v_0+w_{\lambda})}(1-\re^{G(v_0+w_{\lambda})})\ge\frac{4\pi N}{|\Omega|}.\]
Integrating  the above inequality over $\Omega$, we have
\[\lambda\ito \re^{G(v_0+w_{\lambda})}\ud x\ge \lambda \ito \re^{2G(v_0+w_{\lambda})}\ud x+4\pi N>4\pi N,\]
which leads to
  \be
  \lambda \ito \re^{G(v_0+w_{\lambda})}\ud x>4\pi N. \label{b034}
  \ee

 Noting that the function  $G(t)$ is an increasing function which  maps $(-\infty,
 0]$ to itself with
 \[\lim\limits_{t\to-\infty}G(t)=-\infty.\]
 Then we have
 \[\lim\limits_{t\to -\infty}\frac{G(t)}{t}=\lim\limits_{t\to-\infty}G'(t)= \lim\limits_{t\to-\infty}\frac{1}{1-\re^{G(t)}}=1\]
Hence, there exists a positive constant $M$ such that
 \be
 G(t)\le t+1\quad \text{as}\quad t<-M.\label{b035}
 \ee
 Since $v_0+w_\lambda<0$ in $\Omega$,  we decompose $\Omega$ as
  \[\Omega=\Omega_1\cup\Omega_2,\]
where
   \be \Omega_1=\{x\in \Omega|\, v_0+w_\lambda<-M\},\quad\Omega_2=\{x\in \Omega|\, -M\le
   v_0+w_\lambda<0\}.\label{b036}\ee

Hence, by \eqref{b035},  \eqref{b036}, the  the  H\"{o}lder
inequality,  and Trudinger--Moser inequality (see \cite{aubi}),
 \[ \ito \re^{v'}\ud x \le C_1\ito \re^{C_2\|\nabla v'\|^2_2}\ud x, \quad \forall\, v'\in X,  \]
  where $C_1$ and $C_2$ are positive constants, we obtain
 \ber
  \ito \re^{G(v_0+w_{\lambda})}\ud x&=&\int_{\Omega_1} \re^{G(v_0+w_{\lambda})}\ud x+\int_{\Omega_2} \re^{G(v_0+w_{\lambda})}\ud x\nm\\
  &\le& \int_{\Omega_1} \re^{v_0+w_{\lambda}+1}\ud x+|\Omega|\nm\\
  &\le&\re\re^{c_\lambda}\ito \re^{v_0+w'_{\lambda}}\ud x+|\Omega|\nm\\
   &\le&\re\re^{c_\lambda}\left(\ito \re^{2v_0}\ud x\right)^{\frac12} \left(\ito \re^{2w'_\lambda}\ud x\right)^{\frac12}  +|\Omega|\nm\\
   &\le&C_1\re^{c_\lambda}\exp({C_2\|\nabla w'_\lambda\|_2^2})+|\Omega|. \label{b037}
 \eer
  Then from  \eqref{b034},
 \eqref{b037},   and \eqref{b.14},  we obtain a lower bound for  $c_\lambda$,
  \be
  c_\lambda\ge -C(1+\lambda+\lambda^2).\label{b.18}
  \ee

  Consequently,  \eqref{b.15} follows from  \eqref{b.17} and \eqref{b.18}.
  Combining \eqref{b.14}, and \eqref{b.15}, we obtain \eqref{b.16}.

\begin{lemma}\label{lem6}
The set of $\lambda$ for which the equation \eqref{b8} has a
solution is a closed interval. In other words,  at
$\lambda=\lambda_c$,  \eqref{b8} has a solution as well.
\end{lemma}

{\bf Proof.} \quad  For $\lambda_c<\lambda<\lambda_c+1$ (say), by
Lemma \ref{lem5},  the set $\{w_\lambda\}$ is uniformly bounded in
$W^{1,2}(\Omega)$.   Noting the monotonicity of
 $\{w_\lambda\}$   with respect to $\lambda$ in Lemma \ref{lem4}, we conclude
that there exist  a function $\tilde{w}\in W^{1,2}(\Omega)$ such
that
 \[w_\lambda\to \tilde{w }   \quad \text{weakly in }\quad W^{1,2}(\Omega)\quad \text{as} \quad \lambda\to\lambda_c, \]
 and
  \be
  v_0+\tilde{w}< 0  \quad \text{in }\quad \Omega.  \label{b039}
 \ee
Therefore $w_\lambda\to \tilde{w}$ strongly in $L^p(\Omega)$ for any
$p\ge1$ as $\lambda\to\lambda_c$.

  Define
  \[g(t)=\re^{G(t)}\big(\re^{G(t)}-1\big)^2, \quad t\in (-\infty, 0].\]
It is easy to see that
  \[g'(t)=-\re^{G(t)}\big(3\re^{G(t)}-1\big).\]
Since $G(t)<0 $ for all $t<0$, we have
 \be
  |g'(t)|\le 2\quad \text{for all} \quad  t<0. \label{b040}
 \ee
 Hence, in view of   $v_0+w_\lambda<0$, \eqref{b039}, \eqref{b040}, and the fact that $w_\lambda\to \tilde{w}$ strongly in $L^p(\Omega)$ for any
$p\ge1$ as $\lambda\to\lambda_c$,  we  infer  that
 \[  \re^{G(v_0+w_{\lambda})}(\re^{G(v_0+w_{\lambda})}-1)^2 \quad \text{ converges to }\quad \re^{G(v_0+\tilde{w})}(\re^{G(v_0+\tilde{w})}-1)^2, \]
strongly in $L^p(\Omega)$ for any $p\ge1$ as $\lambda\to\lambda_c$.
Using this result in \eqref{b8} and the elliptic $L^2$-estimates, we
 see that  $\tilde{w}\in W^{2,2}(\Omega)$ and $w_\lambda\to \tilde{w}$
strongly in $W^{2, 2}(\Omega)$ as $\lambda\to\lambda_c$.
Particularly, taking the limit $\lambda\to\lambda_c$ in \eqref{b8},
we obtain that $\tilde{w}$ is a solution of \eqref{b8} for
 $\lambda=\lambda_c$.

Finally we show the last statement of Theorem \ref{th2}.

Denote
\[S=\{p_1, \cdots, p_m; n_1, n_2, \cdots, n_m\},\qquad S'=\{p_1, \cdots, p_m; n'_1, n'_2, \cdots, n'_m\}.\]
 We denote the dependence of $\lambda_c$ on $S$ by $\lambda_c(S)$.
 Consider the equation
 \be
  (1-\re^u)\Delta u-\re^u|\nabla u|^2=-\lambda \re^u(\re^u-1)^2+4\pi\sum\limits_{j=1}^mn_j\delta_{p_j} \quad \text{in }\quad \Omega. \label{b20}
 \ee
As before, setting  $v=F(u)=1+u-\re^u$, the equation \eqref{b20} is
equivalent to

 \be
  \Delta v=-\lambda\re^{G(v)}\big(\re^{G(v)}-1\big)^2+4\pi\sum\limits_{j=1}^mn_j\delta_{p_j}\quad \text{in }\quad \Omega.\label{b21}
 \ee

\begin{lemma}\label{lem7}
 If $S\le S'$,  then  $\lambda(S)\le \lambda(S')$.
\end{lemma}

 {\bf Proof.}\quad It is sufficient to prove  that, if $\lambda>\lambda_c(S')$, then $\lambda\ge\lambda_c(S)$.
 Let $v'$ be a solution of \eqref{b21} with $n_j=n'_j, \, j=1, \cdots, m$ and $v_0$ satisfy
 \[\Delta v_0=-\frac{4\pi N}{|\Omega|}+4\pi\sum\limits_{j=1}^mn_j\delta_{p_j},\]
where $N=n_1+\cdots+n_m$. Setting $v'=v_0+w\_$,  we have
 \berr
 \Delta w\_&=&-\lambda\re^{G(v_0+w\_)}\big(\re^{G(v_0+w\_)}-1\big)^2+\frac{4\pi N}{|\Omega|}+4\pi\sum\limits_{j=1}^m(n'_j-n_j)\delta_{p_j}\\
  &\le&-\lambda \re^{G(v_0+w\_)}\big(\re^{G(v_0+w\_)}-1\big)^2+\frac{4\pi N}{|\Omega|},
  \eerr
 in the distributional sense,  which
implies in particular that $w\_$ is a subsolution of \eqref{b8} in
the sense of distribution and \eqref{b12} holds pointwise. It is
easy to check that the singularity of $w\_$ is at most of the type
$\ln|x-p_j|$. Hence, the inequality \eqref{b12} still results in the
convergence of the sequence of $\{w_n\}$ to a solution of \eqref{b8}
in any $C^k$-norm. Indeed, by \eqref{b12}, we see that $\{w_n\}$
converges almost everywhere and is bounded in the  $L^2$-norm.
Therefore, the sequence converges in $L^2(\Omega)$.
 Similarly, we see that  the right-hand side of \eqref{b11} also converges in $L^2(\Omega)$.  Then, it follows from  the standard $L^2$-estimate that   the sequence $\{w_n\}$  converges  in
$W^{2,2}(\Omega)$ to a strong solution of \eqref{b8}.
  Therefore, we can get a classical solution  of \eqref{b8}.  By  a bootstrap argument, we can obtain the convergence in any $C^k$-norm.
   Then we have $\lambda\ge\lambda_c(S)$.    Therefore, $\lambda(S)\le \lambda(S')$.

Then, Theorem \ref{th2} follows from  Lemmas
\ref{le1}$\sim$\ref{lem7}.

\section{Generalized Abelian Higgs vortices }
\setcounter{equation}{0} \setcounter{theorem}{0}

In this section,   we construct a  multivortex solution for the
generalized self-dual Abelian Higgs equation also proposed in
\cite{bct} over the doubly periodic domain $\Omega$, using our
results of the last section.

Recall that in \cite{bct} the  Hamiltonian  of  the generalized
Abelian Higgs model can be written as
 \berr
  H=\ito\ud x\left\{\kappa^2\left[\big(1-|\phi|^2\big)F_{12}-\ri \big(D_1\phi\overline{D_2\phi}-\overline{D_1\phi}D_2\phi\big)\right]^2
\right.\\
\left.+2\big(1-|\phi|^2\big)^2\big(|D_1\phi|^2+|D_2\phi|^2\big)+V\right\}.
 \eerr
 With the choice of the Higgs potential
  \[V=\frac{1}{4\kappa^2}\big(1-|\phi|^2\big)^4, \]
  as in section 2,  we rewrite $H$  as
  \berr
  H&=&\ito\ud x\left\{\left(\kappa\left[\big(1-|\phi|^2\big)F_{12}-\ri
  \big(D_1\phi\overline{D_2\phi}-\overline{D_1\phi}D_2\phi\big)\right]-\frac{1}{2\kappa}\big(1-|\phi|^2\big)^2\right)^2
\right.\\
 &&\left. +2\big(1-|\phi|^2\big)^2|D_1\phi-\ri D_2\phi|^2+\big(1-|\phi|^2\big)^3F_{12}-3\ri\big(1-|\phi|^2\big)^2\big(D_1\phi\overline{D_2\phi}-\overline{D_1\phi}D_2\phi\big)\right\}
\\
 &=&\ito\ud x\left\{\left(\kappa\left[\big(1-|\phi|^2\big)F_{12}-\ri \big(D_1\phi\overline{D_2\phi}-\overline{D_1\phi}D_2\phi\big)\right]
 -\frac{1}{2\kappa}\big(1-|\phi|^2\big)^2 \right)^2\right.\\
 &&\left.+2\big(1-|\phi|^2\big)^2|D_1\phi-\ri D_2\phi|^2+ F_{12}-3\ri\vep_{ij}\partial_i\left[\left(1-|\phi|^2+\frac13|\phi|^4\right)\phi\overline{D_j\phi}\right]\right\}.
 \eerr
Then we obtain
\[
 H\ge \ito F_{12}\ud x,
 \]
 and this  lower bound is saturated if and only if  $(\phi, A)$ satisfies  the self-dual
 equations
  \ber
   D_1\phi&=&\ri D_2\phi, \label{c1} \\
  F_{12}&=&\ri\big(D_1\phi\overline{D_2\phi}-\overline{D_1\phi}D_2\phi\big)+\frac{1}{2\kappa^2}\big(1-|\phi|^2\big)^2.\label{c2}
  \eer
 The structure of \eqref{c1} and \eqref{c2} is  similar to that of
 \eqref{a1} and \eqref{a2}. However, the approach dealing with  \eqref{a1} and
 \eqref{a2} cannot be directly used to  \eqref{c1} and \eqref{c2}.
 Fortunately, based on the obtained solution of \eqref{a1} and \eqref{a2}, we
 we can establish a solution of \eqref{c1} and \eqref{c2}.

  Following a similar procedure as in section 2,  we can reduce the equations \eqref{c1} and
  \eqref{c2} into the quasilinear elliptic equation
 \be
  \Delta u-\re^u|\nabla u|^2=-\lambda  (\re^u-1)^2+4\pi\sum\limits_{s=1}^N\delta_{p_s} \quad \text{in }\quad \Omega, \label{c3}
 \ee
 where  $\lambda=\frac{1}{\kappa^2}$.
 Using  $v=F(u)=1+u-\re^u$ again, we have
   \be
    \Delta v=-\lambda\big(\re^{G(v)}-1\big)^2+4\pi\sum\limits_{s=1}^N\delta_{p_s} \quad \text{in }\quad \Omega.\label{c4}
   \ee
 Let $v=v_0+w$, where $v_0$ is defined by \eqref{b7}. Then the equation \eqref{c4}  is modified into
 \be
  \Delta w=-\lambda \big(\re^{G(v_0+w)}-1\big)^2+ \frac{4\pi N}{|\Omega|} \quad \text{in }\quad
  \Omega.\label{c5}
 \ee

 Let $\underline{w}$ be a solution of \eqref{b8}. Then we have
 $v_0+\underline{w}<0$ in $\Omega$. As a result,
 $\re^{G(v_0+\underline{w})}<1,$  which implies
  \berr
  \Delta \underline{w}&=&-\lambda\re^{G(v_0+\underline{w})}\big(\re^{G(v_0+\underline{w})}-1\big)^2 + \frac{4\pi N}{|\Omega|}\\
   &\ge&-\lambda\big(\re^{G(v_0+\underline{w})}-1\big)^2+ \frac{4\pi N}{|\Omega|}.
 \eerr
Thus  we see that $\underline{w}$ is a subsolution of \eqref{c5}. It
is easy to see that $-v_0$ is also a supersolution  of \eqref{c5}.
Therefore we can modify the iteration scheme \eqref{b11} to
establish a solution $w$ of \eqref{c5}, satisfying
$\underline{w}<w<-v_0$. Indeed, we can get the following theorem.
\begin{theorem}\label{th3}
 For any prescribed  points $p_1, \dots, p_N\in\Omega$,  there is a critical value of $\lambda$, say  $\lambda_c$, satisfying
 \[
  \lambda_c\ge \frac{4\pi N}{|\Omega|},
 \]
   such that, for $\lambda>\lambda_c,$ the equation \eqref{c3} has a   solution, while for $\lambda\le\lambda_c$,
  the equation  \eqref{c3} has no solution.
\end{theorem}
\begin{remark}
  It was shown in  \cite{waya},  for the  Abelian Higgs   equation,
  \be
  \Delta u=\lambda(\re^u-1)+4\pi\sum\limits_{j=1}^N\delta_{p_j}  \quad \text{in }\quad \Omega,
  \label{c6}
  \ee
  the critical value of the coupling parameter is
  \[\lambda_c=\frac{4\pi N}{|\Omega|}.\]
  But, at $\lambda=\lambda_c$, the equation \eqref{c6} has no
  solution.
\end{remark}
Consequently, by Theorem \ref{th3}, we can recover a solution to
\eqref{c1} and \eqref{c2} by the transformation \eqref{b4'} and
\eqref{b4''}.
 \begin{theorem} \label{th4}
   Let $p_1, p_2, \dots, p_m\in\Omega$, $n_1, n_2, \dots, n_m$ be
   some  positive integers and $N=\sum_{i=1}^mn_i$. There exists a critical
   value of the coupling parameter, say $\kappa_c$, satisfying
   \[0<\kappa_c \le  \frac12\sqrt{\frac{|\Omega|}{\pi N}},\]
   such that, for $0<\kappa<\kappa_c$,  the self-dual equations
 \eqref{c1} and \eqref{c2} admit a solution $(\phi, A)$ for which $p_1,
 p_2,\dots,p_m$ are zeros of $\phi$ with multiplicities $n_1, n_2, \dots,
 n_m$,  while for  $ \kappa\ge\kappa_c$, the equations  \eqref{c1} and \eqref{c2} have no
solution.
  \end{theorem}

\end{document}